\documentclass[a4paper,12pt]{article}
\usepackage[english]{babel}
\usepackage[numbers,comma,square,compress]{natbib}
\usepackage{amssymb,amsmath,amsfonts,bm,braket,enumerate}

\usepackage{dcolumn}
\usepackage{bm}
\usepackage{ifpdf}
\usepackage{xcolor,color,graphicx,graphics}
\usepackage[OT1]{fontenc}
\usepackage{latexsym}
\usepackage{makeidx}
\usepackage{epsfig,subfigure}
\usepackage{epstopdf}
\usepackage{mathrsfs}
\usepackage{enumerate}

\usepackage[colorlinks=true,
            linkcolor=blue,
            urlcolor=blue,
            citecolor=blue]{hyperref}

\makeatletter\renewcommand\section{\@startsection {section}{1}{\z@}%
                                   {-3.5ex \@plus -1ex \@minus -.2ex}
                                   {2.3ex \@plus.2ex}%
                                   {\normalfont\large\bfseries}}
\renewcommand\subsection{\@startsection{subsection}{2}{\z@}%
                                     {-3.25ex\@plus -1ex \@minus -.2ex}%
                                     {1.5ex \@plus .2ex}%
                                     {\normalfont\bfseries}}

\everymath{\displaystyle}

\newcommand{\bea}{\begin{eqnarray}}
\newcommand{\eea}{\end{eqnarray}}

\newcommand{\email}[1]{\footnote{E-mail: \href{mailto:#1}{#1}}}
\newcommand{\orcid}[1]{\href{https://orcid.org/#1}{\includegraphics[width=10pt]{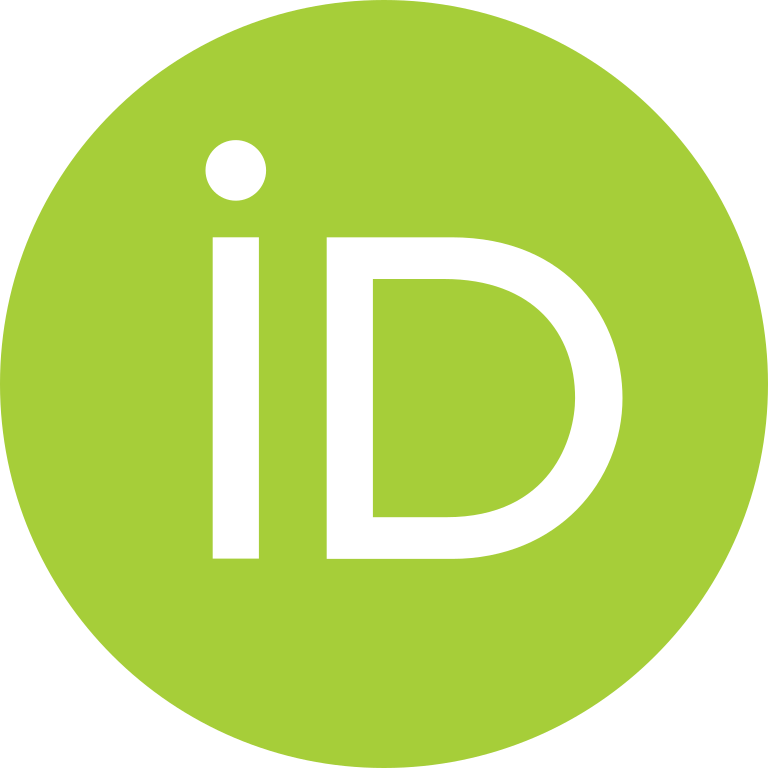}}}

\usepackage{fixmath}
\usepackage{mathtools,slashed}

\begin{document}

\title{Vacuum Cherenkov radiation at finite temperature}


\author{R. Bufalo$^{1}$ \orcid{0000-0003-1879-1560}\email{rodrigo.bufalo@ufla.br} ~and  A. F. Santos$^{2}$ \orcid{0000-0002-2505-5273} \email{alesandroferreira@fisica.ufmt.br} \\
\textit{$^{1}$ \small Departamento de F\'isica, Universidade Federal de Lavras,}\\
\textit{ \small Caixa Postal 3037, 37200-900 Lavras, MG, Brazil}\\
\textit{$^{2}$ \small Instituto de F\'{\i}sica, Universidade Federal de Mato Grosso,}\\
\textit{ \small 78060-900, Cuiab\'{a}, Mato Grosso, Brazil}
}

\maketitle
\date{}

\begin{abstract}
In this paper we examine the thermal effects of the vacuum Cherenkov radiation in a Lorentz- and CPT-violating electrodynamics.
We compute the thermal contribution to the Cherenkov radiation rate within the Thermofield Dynamics approach.
Since the model under consideration possess a consistent canonical quantization and  also fulfils the physical constraints in order to this vacuum process to happen, it is a perfect candidate to implement the study at finite temperature.
We evaluate in details the instantaneous rate of energy loss for a charge, and show that the radiation rate is significantly modified at very high temperatures.
Intriguingly, we further observe that when the temperature goes to infinity the radiation rate goes to zero even if the process is kinematically allowed.
\end{abstract}

\section{Introduction}

Models involving Lorentz violation have reached an important milestone in recent years due to its systematic development and to the increasing number of precision tests they have been subject \cite{ref53,AmelinoCamelia:2008qg}.
A richer context offering valuable prospects about departures from Lorentz symmetry, and making contact with the so-called Physics Beyond the Standard Model, is the anomalous decay processes  \cite{Lykken:2010mc}.
Since they are affected in unexpected ways by Lorentz violation, it might happens that forbidden processes can occur in certain regions of the parameter space \cite{ref53,Jacobson:2005bg}.

In this scenario, models involving instabilities of photons in vacuum have caught interest in recent years because sufficiently energetic photon (usually from gamma-ray bursts) may decay as a manifestation of Lorentz  violation \cite{Ellis:2005wr,Heeck:2013cfa,Bonetti:2017pym}.
These instabilities of highly energetic particles are, in general, related to manifestations of Lorentz violation.
A phenomenologically important anomalous decay process, in the framework of Lorentz violating photons, is the emission of vacuum Cherenkov radiation, a significant energy loss process for high-energy particles \cite{Lehnert,Kaufhold:2005vj,Kaufhold:2007qd}.

Many aspects about the possibility of vacuum Cherenkov radiation
by Lorentz violating effects have been extensively discussed in the framework of the Standard-Model Extension (SME), within the classical approach to electromagnetic particle radiation \cite{Altschul:2006zz,Altschul:2007tn,Altschul:2007kr,Schober:2015rya,DeCosta:2018nyf}, as well as in the field theory \cite{Kaufhold:2005vj,Kaufhold:2007qd,Colladay:2016rmy,Schreck:2017isa} and also in the Lifshitz-like electrodynamics   \cite{Bufalo:2021myy}.
However, to the best of our knowledge no study of vacuum Cherenkov radiation has been performed in the presence of a thermal bath.
Since it is well known that physical systems behavior can drastically change due to the presence of thermal effects, one can naturally asks in this context: can temperature effects change significantly the emission of Cherenkov radiation in vacuum, by enhancing the radiation rate or even by prohibiting it to happen?
In order to address these and other questions we will approach the vacuum Cherenkov radiation in terms of the Thermofield Dynamics formalism.

Along Matsubara's imaginary time formalism and Schwinger-Keldysh closed time path method \cite{Matsubara, Schwinger, kapusta}, the thermofield dynamics formalism \cite{Takahashi,Umezawa:1982nv,Umezawa:1993yq,Khanna:2009zz} is one of the most important approaches to describe finite temperature field theory.
The key feature of the thermofield dynamics method is that the Fock space is doubled, then you have a new set of operators, designed tilde operators, acting on the second Fock space, tilde space.
Physically speaking, the second Fock space is interpreted as a heat bath that ensures the dynamical system to stay in equilibrium.  
In addition to the duplicated space, another fundamental element of this formalism is the Bogoliubov transformation. This transformation consists of a rotation between two spaces, original and tilde, which introduces the thermal effects.
On the algebraic aspect, the most appealing advantage of the thermofield dynamics method is that you can carry out analysis and calculations of scattering amplitudes and decay processes exactly as in the case of $T=0$ field theory.

Hence, since finite temperature effects have profound implications in the study of high energy physics,  we will examine the behavior of the vacuum Cherenkov radiation within a well known CPT- and Lorentz violating electrodynamics in the presence of a heat bath.
We start Sec.~\ref{sec2} by reviewing the main aspects and definition of the CPT- violating electrodynamics, where the fermionic sector is the standard Dirac Lagrangian, while the photon sector is modified by terms belonging to the minimal sector of the SME.
Moreover, we establishes the modified dispersion relations and polarization states for the photon field.
We discuss in Sec.~\ref{sec3} the features of the vacuum process $e^- \to \gamma + e^-$ at finite temperature.
First, we present in details the calculation of the transition amplitude at finite temperature related with the vacuum Cherenkov decay within the thermofield dynamics framework.
Furthermore, we compute the rate of radiated energy at the high-temperature regime and analyse how the thermal effects changes the $T=0$ results. In particular, we discuss how the energy of the thermal bath can prohibits the Cherenkov decay to happen even if it is kinematically allowed.
In Sec.\ref{conc} we summarize the results, and present our final remarks.

\section{CPT-violating electrodynamics}
\label{sec2}

It is well known that ordinary Cherenkov radiation can only occur for particles propagating in a medium, since a Lorentz-invariant vacuum prevents it by energy-momentum conservation \cite{jelley,Macleod:2018zcb}.
However, some Lorentz violating scenarios provide sufficient instabilities so that particles can radiate through the Cherenkov process even in vacuum \cite{Lehnert,Altschul:2017xzx}.

In this context, we choose to conduct our analysis at finite temperature in a well-behaved model where this vacuum decay process is known to happen at zero temperature.
Moreover, we focus in modifying only the radiation sector with a minimal deformation and preserving the matter dynamics. 
Hence, we shall review in this section the main aspects involving the following Lorentz- and CPT-violating Lagrangian density \cite{Colladay:2016rmy}
\begin{equation}
{\cal L}={\cal L}_\psi+{\cal L}_A,
\end{equation}
with the standard fermionic field Lagrangian
\begin{equation}
{\cal L}_\psi=  \bar{\psi}\left(i\slashed{\partial}-ie\slashed{A}+m\right)\psi\label{fermion}
\end{equation}
and
\begin{equation} \label{eqA}
{\cal L}_A=-\frac{1}{4}F_{\mu\nu}F^{\mu\nu}+\frac{1}{2}k_{AF}^\kappa\epsilon_{\kappa\lambda\mu\nu}A^\lambda F^{\mu\nu}+\frac{1}{2}m_\gamma^2A_\mu A^\mu-\frac{1}{2\xi}(\partial_\mu A^\mu)^2
\end{equation}
is the photon Lagrangian that belongs to the minimal sector of the SME. Here $k_{AF}^\kappa$ is an arbitrary fixed background vector, $m_\gamma$ is the photon mass and $\xi$ is a gauge parameter. 

There are a number of well known studies of the vacuum Cherenkov radiation for the model \eqref{eqA} for the case of spacelike $k_{AF}^\kappa$ 
\cite{Lehnert,Kaufhold:2007qd} and also for purely timelike $k_{AF}^\kappa$ \cite{Colladay:2016rmy}. We shall focus our analysis at finite temperature for the timelike case, and present some remarks about the spacelike model.

It is important to emphasize that the presence of the photon's mass $m_\gamma$ in \eqref{eqA} is related with the fact that for timelike $k_{AF}^\kappa$ the photon's dispersion relation is tachyonic (leading to an unstable theory), and this (small) mass term circumvents this problem \cite{Alfaro:2006dd,Colladay:2016rmy}.
It is important to stress that this mechanism does not violate any experimental observation \cite{Colladay:2016rmy}.
Hence, from the field equations the full dispersion relation for the photon is cast as
\begin{equation} \label{disp_rel}
\left(p^{2}-\xi m_{\gamma}^{2}\right)\left(p^{2}-m_{\gamma}^{2}\right)\left[\left(p^{2}-m_{\gamma}^{2}\right)^{2}+4\left(k_{AF}^{2}p^{2}-\left(k_{AF}\cdot p\right)^{2}\right)\right]=0.
\end{equation}
From this expression we can immediately determine the polarization states $\epsilon_{(\alpha)}^{\lambda}(\vec{p})$, which we assume to depend on the spacelike three-momentum.
The first ones we consider are
\begin{equation}
\epsilon^{\left(0\right)\mu}\left(\vec{p}\right)=N_{0}p^{\mu},\quad\epsilon^{\left(3\right)\mu}\left(\vec{p}\right)=N_{3}\left(k_{AF}^{\mu}-\frac{\left(p.k_{AF}\right)}{m_{\gamma}^{2}}p^{\mu}\right),
\end{equation}
where $N_{0,3}$ are normalization constants, and we have assumed $m_\gamma> 0$.
Now, under the consideration that $k^{\mu}_{AF}$ is timelike ($\vec{k}_{AF}=0$ and $k^0_{AF}> 0$ ), the modified modes are given by
\begin{equation} \label{pert_dr}
\omega_{p}^{\left(\pm\right)2}=\left|\vec{p}\right|^{2}+\mu_{\pm}^{2},
\end{equation}
where we have defined
\begin{equation}
\mu_{\pm}^{2}=m_{\gamma}^{2}\pm2k_{AF}^{0}\left|\vec{p}\right|.
\end{equation}
We can observe from eq.~\eqref{pert_dr} that in order to always have $\omega_{p}^{\left(\pm\right)2}>0$ it is necessary that $m_\gamma \geq k^0_{AF}$ (see that problems can arise when $m_\gamma \sim k^0_{AF}$).
One solution to eq.~\eqref{pert_dr} reads
\begin{equation} \label{remaining_dr}
\epsilon^{\left(\pm\right)}\left(\vec{p}\right)=N_{\pm}\left(\begin{array}{c}
0\\
p^{1}p^{2}\mp ip^{3}\left|\vec{p}\right|\\
-p_{1}^{2}-p_{3}^{2}\\
p^{2}p^{3}\mp ip^{1}\left|\vec{p}\right|
\end{array}\right),
\end{equation}
where $N_{\pm}$ are normalization constants, and that $k_{AF}\cdot k_{AF}>0 $ and $\vec{k}_{AF}=0$.
These physical modes $\epsilon^{\left(\pm\right)}$ shown in eq.~\eqref{remaining_dr} are in general spacelike and exhibit birefringent behavior (crucial for the vacuum radiation to occur, see discussion below).

As a matter of fact, the dynamical constraint for the Cherenkov radiation to happens requires that the fermion group velocity $v_g$ should be greater than the photon phase velocity $v_{\rm ph}$ for some values of the three-momentum  in medium materials \cite{jelley}, but also for Lorentz-violating vacua \cite{Altschul:2006zz,Altschul:2007tn,Altschul:2007kr}.
This can only be achieved in this model by the state $\epsilon^{\left(-\right)}$, because that for sufficiently large three-momentum the photon four-momentum is spacelike (see eq.~\eqref{pert_dr}) and therefore $v_{\rm ph}$ is subluminal, satisfying the above criteria. 

Before proceeding to the evaluation of the thermal contribution to the Cherenkov radiation rate we shall review in the next section some important aspects of the thermofield dynamics formalism useful for our development.


\section{Cherenkov process at finite temperature}
\label{sec3}

The quantum decay process related with the vacuum Cherenkov radiation is described by	
\bea
e^-(q,s_i)\rightarrow e^-(q',\lambda)+\gamma(p,s_f),
\eea
which corresponds to an initial particle with momentum and spin $(q,s_i)$, decays into a particle $(q',s_f)$ and emits a Cherenkov radiation $(p,\lambda)$, where $(s_i,s_f)$ are the spin of the initial and final particles, respectively, while $\lambda$ is the polarization of the radiation. 

Since our main objective is to evaluate the energy loss associated with the vacuum Cherenkov scattering at finite temperature, let us review some general results necessary for the development of this analysis \cite{Kaufhold:2005vj,Kaufhold:2007qd,Bufalo:2021myy}.
The energy-momentum loss of a Lorentz invariant charged particle per unit of time is equal to the photon four-momentum $p$ weighted by the scattering amplitude squared and integrated over the phase space,
\begin{equation}
\frac{dq^{\mu}}{dt}=\int Dp\left|\mathcal{M}\right|^{2}p^{\mu}
\end{equation}
where $Dp$ is the phase-space invariant measure.
The rate of total radiated energy is obtained from the time
component of the above expression.

Thus, in the leading order at the CPT- and Lorentz-violating parameter $k_{AF}^{0}$, the rate of radiated energy can be expressed as
\begin{equation}
W\approx-\dot{q}_{0} = \int p_{0}d\Gamma, \label{eq100}
\end{equation}
where $d\Gamma$ is the differential decay rate for the given process and can readily be obtained \cite{mandl_shaw}
\bea \label{eq166}
d\Gamma(\beta)=(2\pi)^4\frac{1}{2q^0}4m^2\frac{d^3\vec{p}}{(2\pi)^3(p^0)}\frac{d^3\vec{q'}}{(2\pi)^3(q'^0)}\delta^4(q'+p-q)\frac{1}{2}\sum_{\rm spins}|{\cal M}(\beta)|^2.
\eea
Here the normalization factors are chosen accordingly for bosonic
and fermionic fields, and we have averaged initial spins and summer over final spins.
Observe in eq.~\eqref{eq166} that the temperature effects are solely contained in the transition amplitude ${\cal M}(\beta)$.

Since we wish to evaluate the radiation rate through the differential decay rate (see eq.~\eqref{eq100}), the main object to compute in this analysis is the transition amplitude related with  the vacuum decay process $e^- \to e^-+\gamma$.
We shall review, in the next section, the main aspects of the thermofield dynamics approach for the evaluation of the transition amplitude \cite{Umezawa:1982nv,Umezawa:1993yq,Khanna:2009zz}.

\subsection{Thermofield dynamics}
\label{sec4}

We formally define the transition amplitude at finite temperature ${\cal M}(\beta)$ as
\bea
{\cal M}(\beta)=\langle f,\beta| \, \hat{S} \,| i,\beta\rangle,
\eea
where the $\hat{S}$-matrix is given
\bea \label{eq200}
\hat{S}=\sum_{n=0}^\infty\frac{(-i)^n}{n!}\int dx_1dx_2\cdots dx_n \tau \left[ \hat{{\cal L}}_{I}(x_1) \hat{{\cal L}}_{I}(x_2)\cdots \hat{{\cal L}}_{I}(x_n) \right],
\eea
with $\tau$ being the time ordering operator and
\bea
\hat{{\cal L}}_{I}(x)={{\cal L}}_{I}(x)-\tilde{{\cal L}}_{I}(x)
\eea
describes the Lagrangian interaction part in the doubled notation of the TFD formalism.
In this approach, the thermal states of the process $e^-(q)\rightarrow e^-(q')+\gamma(p)$ are defined as
\begin{align}
| i,\beta\rangle =\, &c_{q,s_i}^\dagger(\beta)|0(\beta)\rangle, \cr 
 | f,\beta\rangle = \,&a_p^\dagger(\beta)c_{q',s_f}^\dagger(\beta)|0(\beta)\rangle ,
\end{align}
with $c_q^\dagger(\beta)$ and $a_{p}^\dagger(\beta)$ being the creation operators for fermions and bosons, respectively. 

At the tree level, the $\hat{S}$-matrix \eqref{eq200} becomes
\bea
\hat{S}=-i\int d^4x\,\left({{\cal L}}_{I}(x)-\tilde{{\cal L}}_{I}(x)\right),
\eea
thus the transition amplitude is given as
\bea \label{eq_M}
{\cal M}(\beta)=-i\int d^4x\,\Bigl\langle 0(\beta)\Bigl |a_p(\beta)c_{q',s_f}(\beta)\Bigl({\cal L}_I(x)-\tilde{\cal L}_I(x)\Bigl)c_{q,s_i}^\dagger(\beta)\Bigl | 0(\beta)\Bigl\rangle.
\eea
One can read from the Lagrangian density eq.~\eqref{fermion} the photon-matter coupling
\begin{align}
{\cal L} (x)&=-ie\bar{\psi}\gamma^\mu\psi A_\mu,\\
\tilde{{\cal L}}(x)&= -ie\tilde{\bar{\psi}}\gamma^\mu\tilde{\psi}\tilde{ A_\mu}.
\end{align}
Hence, the amplitude \eqref{eq_M} can be written in a simple fashion as
\bea
{\cal M}(\beta)=-e \int d^4x\,{\cal M}_1(\beta){\cal M}_2(\beta)+e \int d^4x\,\widetilde{{\cal M}}_1(\beta)\widetilde{{\cal M}}_2(\beta), \label{TA}
\eea
with the definitions
\begin{align}
{\cal M}_1(\beta) = \,&\Bigl\langle 0(\beta)\Bigl |\psi(x) c_{q,s_i}^\dagger(\beta)\Bigl | 0(\beta)\Bigl\rangle, \label{M1}\\
{\cal M}_2(\beta)=\, &\Bigl\langle 0(\beta)\Bigl |a_p(\beta)c_{q',s_f}(\beta)\bar{\psi}(x)\gamma^\mu A_\mu(x)\Bigl | 0(\beta)\Bigl\rangle,  \label{M2}\\
\widetilde{{\cal M}}_1(\beta)=\,&\Bigl\langle 0(\beta)\Bigl |\widetilde{\bar{\psi}}(x) c_{q,s_i}^\dagger(\beta)\Bigl | 0(\beta)\Bigl\rangle, \label{M11}\\
\widetilde{{\cal M}}_2(\beta)=\,&\Bigl\langle 0(\beta)\Bigl |a_p(\beta)c_{q',s_f}(\beta)\widetilde{\psi}(x)\gamma^\mu \widetilde{A}_\mu(x)\Bigl | 0(\beta)\Bigl\rangle. \label{M22}
\end{align}
It is interesting to remark that eq.~\eqref{TA} displays explicitly the doubleness of the Fock space related with the TFD approach.

In order to calculate these matrix elements separately, we shall consider the (doubled) fermion field written as
\bea
\psi(x)&=&\sum_s\int \frac{d^3p}{(2\pi)^{3/2}}\, N_p\left[c_{p,s} u_s(p)e^{-ipx}+d_{p,s}^\dagger v_s(p)e^{ipx}\right],\\
\tilde{\psi}(x)&=&\sum_s\int \frac{d^3p}{(2\pi)^{3/2}}\, N_p\left[\tilde{c}_{p,s} \tilde{u}_s(p)e^{ipx}+\tilde{d}_{p,s}^\dagger \tilde{v}_s(p)e^{-ipx}\right],
\eea
where $N_p$ is the normalization constant and $u_\alpha(p)$ and $v_\alpha(p)$ are Dirac spinors, and we have the (tilde) conjugation operation $\tilde{u}(p,s)=u^\dagger(p,s) $ \cite{Umezawa:1982nv,Umezawa:1993yq,Khanna:2009zz}. In addition, the photon field in terms of Fourier modes is given by
\bea
A_{\mu}(x)=\int \frac{d^3k}{(2\pi)^3}\sum_\lambda\frac{1}{2p_0^\lambda}\left(a^\lambda(k)\epsilon_\mu^{(\lambda)}(k)e^{-ik\cdot x}+a^{\lambda\dagger}(k)\epsilon_\mu^{*(\lambda)}(k)e^{ik\cdot x}\right),\label{A1}
\eea
with $\epsilon_{\mu}^{(\lambda)}(k)$ being the polarization vector of the physical polarization states.

Using the fermionic field, the first matrix element eq.\eqref{M1}  becomes
\bea
{\cal M}_1(\beta)=\sum_r\int \frac{d^3p}{(2\pi)^{3/2}}\, N_p\Bigl\langle 0(\beta)\Bigl |c_{p,r} c_{q,s_i}^\dagger(\beta)u_r(p)e^{-ipx}\Bigl | 0(\beta)\Bigl\rangle. \label{matrix_M}
\eea
We observe that for fermions, with the operators $c_{p,s}^\dagger$ and $c_{p,s}$ being the creation and annihilation operators, respectively, the Bogoliubov transformations lead to the relations (analogous relations hold for $d$ and $d^\dagger$) \cite{Umezawa:1982nv,Umezawa:1993yq,Khanna:2009zz}
\bea
c_{p,s}&=&\cos\theta_p\, c_{p,s}(\beta)+i\sin\theta_p\,\tilde{c}^\dagger_{p,s}(\beta),\label{c1}\\
c^\dagger_{p,s}&=&\cos\theta_p\, c^\dagger_{p,s}(\beta)-i\sin\theta_p\,\tilde{c}_{p,s}(\beta), \label{c2}\\
\tilde{c}_{p,s}&=&\cos\theta_p\, \tilde{c}_{p,s}(\beta)-i\sin\theta_p\,{c}^\dagger_{p,s}(\beta), \label{c3}\\
\tilde{c}^\dagger_{p,s}&=&\cos\theta_p\, \tilde{c}^\dagger_{p,s}(\beta)+i\sin\theta_p\,{c}_{p,s}(\beta), \label{c4}
\eea
with
\begin{align}
\sin^2\theta_p&=1/(1+e^{\beta p_0 })\equiv n_F(p), \cr
\cos\theta_p&=e^{\beta p_0/2}\sin\theta_p,\label{F}
\end{align}
where $n_F(p)$ corresponds to the Fermi-Dirac distribution, and we have also assumed, by simplicity, the chemical potential to be zero.
The (equal time) anti-commutation relations for the creation and annihilation fermionic operators at finite temperature are given by
\bea
\left\{c_{p,s}(\beta), c^\dagger_{q,r}(\beta)\right\}=\delta_{rs}\delta(\vec{p}-\vec{q}),\quad\quad\quad \left\{\tilde{c}_{p,s}(\beta), \tilde{c}^\dagger_{q,r}( \beta)\right\}=\delta_{rs}\delta(\vec{p}-\vec{q}),\label{ComF}
\eea
and other commutation relations are null.
On the other hand, for bosons, the Bogoliubov transformations are  \cite{Umezawa:1982nv,Umezawa:1993yq,Khanna:2009zz}
\bea
a_p&=&\cosh\theta_p \,a_p(\beta)+\sinh\theta_p\,\tilde a_p^\dagger(\beta),\label{BTp1} \\
a^\dagger_p&=&\cosh\theta_p \,a^\dagger_p(\beta)+\sinh\theta_p\,\tilde a_p(\beta),\label{BTp2} \\
\tilde a_p&=&\cosh\theta_p\,\tilde a_p(\beta)+\sinh\theta_p\, a^\dagger_p(\beta),\label{BTp3}  \\
\tilde a^\dagger_p&=&\cosh\theta_p\,\tilde a^\dagger_p(\beta)+\sinh\theta_p\, a_p(\beta),\label{BTp}
\eea
with
\bea
\sinh^2\theta_p&=&1/(e^{\beta p_0}-1)\equiv n_B(p)\nonumber\\
\cosh\theta_p&=&e^{\beta p_0/2}\sinh\theta_p,\label{B}
\eea
where $n_B(p)$ corresponds to the Bose-Einstein distribution.
The creation and annihilation operators satisfy commutation relations
\bea
\left[a_p(\beta), a^\dagger_q(\beta)\right]=\delta(\vec{p}-\vec{q}),\quad\quad\quad \left[\tilde{a}_p(\beta), \tilde{a}^\dagger_q(\beta)\right]=\delta(\vec{k}-\vec{p}),\label{ComB}
\eea
and other commutation relations are null.

Applying the Bogoliubov transformation \eqref{c1} for $c_{p,s}$ in the matrix element ${\cal M}_1(\beta)$  \eqref{matrix_M}, we get
\bea \label{eq39}
{\cal M}_1(\beta)=\sum_r\int \frac{d^3p}{(2\pi)^{3/2}}\, N_p\Bigl\langle 0(\beta)\Bigl |\cos\theta_p c_{p,r}(\beta) c_{q,s_i}^\dagger(\beta)u_r(p)e^{-ipx}\Bigl | 0(\beta)\Bigl\rangle.
\eea
We can solve eq.~\eqref{eq39} using the anti-commutation relation \eqref{ComF} and performing the momentum integration, so that we obtain
\bea \label{eq_m1}
{\cal M}_1(\beta)=  N_q \cos\theta_q u_{s_i}(q)e^{-iqx}.
\eea

Moreover, we can evaluate the matrix element \eqref{M2} by substituting the fermion and photon fields
\begin{align}
{\cal M}_2(\beta)&=\Bigl\langle 0(\beta)\Bigl |a_p ^{\lambda }(\beta)c_{q',s_f}(\beta)\sum_r\int \frac{d^3k}{(2\pi)^{3/2}}\, N_k\left[c_{k,r}^\dagger \bar{u}_r(k)e^{ikx}+d_{k,s} \bar{v}_s(k)e^{-ikx}\right]\gamma^\mu\nonumber\\
&\times  \int \frac{d^3k'}{(2\pi)^3}\sum_{\lambda'}\frac{1}{2k_0^\lambda}\left(a^{\lambda'} (k')\epsilon_\mu^{(\lambda')}(k')e^{-ik'\cdot x}+a^{\lambda'\dagger}(k')\epsilon_\mu^{*(\lambda')}(k')e^{ik'\cdot x}\right)\Bigl | 0(\beta)\Bigl\rangle.
\end{align}
This expression can be solved by applying the Bogoliubov transformations for fermion and bosons and using their respective commutation relations, resulting in the following expression
\begin{align}
{\cal M}_2(\beta)&=  \int \frac{d^3k}{(2\pi)^{3/2}}\, N_k\int \frac{d^3k'}{(2\pi)^3}\frac{1}{2k_0^{'\lambda}}\nonumber\\
&\times \Bigl\langle 0(\beta)\Bigl |\delta(\vec{k}-\vec{q'})\cos\theta_k \bar{u}_{s_f}(k)e^{ikx} \gamma^\mu\epsilon_\mu^{*(\lambda)}(k')\delta(\vec{k'}-\vec{p})\cosh\theta_{k'}e^{-ik'x}\Bigl | 0(\beta)\Bigl\rangle.
\end{align}
At last, performing the integrals over the momentum variables $k$ and $k'$ we find
\bea \label{eq_m2}
{\cal M}_2(\beta)= \frac{N_{q'}}{2p_0^{\lambda}}\cos\theta_{q'}\cosh\theta_{p} \bar{u}_{s_f}(q')\gamma^\mu\epsilon_\mu^{*(\lambda)}(p)e^{i(q'+p)x}.
\eea

The remaining parts of the amplitude \eqref{TA} can be readily evaluated by performing similar steps as those presented above. Hence, the tilde parts are found as
\bea \label{eq_m3}
\widetilde{{\cal M}}_1(\beta)= i N_q \sin\theta_q \tilde{u}_{s_i}(q)e^{-iqx}
\eea
and 
\bea \label{eq_m4}
\widetilde{{\cal M}}_2(\beta)=-i \frac{N_{q'}}{2p_0^{\lambda}}\sin\theta_{q'}\sinh\theta_{p} \tilde{u}_{s_f}(q')\gamma^\mu\epsilon_\mu^{*(\lambda)}(p)e^{i(q'+p)x}.
\eea

Finally, in terms of the results eqs.~\eqref{eq_m1}, \eqref{eq_m2}, \eqref{eq_m3} and \eqref{eq_m4}, the transition amplitude \eqref{TA} becomes
\begin{align} \label{TA_1}
{\cal M}(\beta)=& \,-e\int d^4x\,e^{i(q'+p-q)x} \frac{N_qN_{q'}}{2p_0^{\lambda}}\cos\theta_q\cos\theta_{q'}\cosh\theta_p\,u_{s_i}(q)\bar{u}_{s_f}(q')\gamma^\mu\epsilon^{*(\lambda)}_\mu(p)\nonumber\\
&+e\int d^4x\,e^{i(q'+p-q)x} \frac{N_qN_{q'}}{2p_0^{\lambda}}\sin\theta_q\sin\theta_{q'}\sinh\theta_p\,\tilde{\bar{u}}_{s_i}(q)\tilde{u}_{s_f}(q')\gamma^\mu\epsilon^{*(\lambda)}_\mu(p).
\end{align}
The integration over $x$ expresses overall four-momentum conservation
\bea
\int d^4x e^{-i(q-q'-p)x}=\delta^4(q-q'-p),
\eea
which will be omitted in the calculations that follow.
Hence, the expression \eqref{TA_1} is simply written as follows
\bea
{\cal M}(\beta)=-  \frac{e N_q N_{q'}}{2p_0^{\lambda}}\gamma^\mu\epsilon^{*(\lambda)}_\mu(p)\Bigl[\cos\theta_q\cos\theta_{q'}\cosh\theta_p
-\sin\theta_q\sin\theta_{q'}\sinh\theta_p\Bigl]\,u_s(q)\bar{u}_s(q'),\label{TA2}
\eea
where we have made use of the tilde conjugation operation mentioned before.
One can observe that the expression \eqref{TA2} has the same structure of the usual one at $T=0$, and that the temperature effects are contained in the trigonometric factors. These temperature factors will be easily expressed in terms of the Fermi-Dirac or Bose-Einstein distributions in the following analysis.  

In order to calculate the differential decay rate eq.~\eqref{eq166}, the main quantity that must be calculated is the squared of the transition amplitude,
\bea
\frac{1}{2}\sum_{\rm spins}|{\cal M}(\beta)|^2=\frac{1}{2}\sum_{\rm spins}{\cal M}(\beta){\cal M}^*(\beta).
\eea
Moreover, in terms of the result \eqref{TA2} and also the completeness relation $\sum_r u_r(p)\bar{u}_r(p)= \frac{\gamma.p+m}{2m}$, the last equation takes the form
\begin{align} \label{eq50}
\frac{1}{2}\sum_{\rm spins}|{\cal M}(\beta)|^2=&\,\frac{e^2}{8}\frac{m^2}{ \omega_p^{(-)2}E_q E_{q-p}} \mathrm{tr}\left[\frac{\gamma.q+m}{2m}\gamma^\mu\epsilon^{*(-)}_\mu(p)\frac{\gamma. (q-p)+m}{2m}\gamma^\nu\epsilon^{(-)}_\nu(p) \right]\nonumber\\
&\times \left[\cos\theta_q\cos\theta_{q'}\cosh\theta_p-\sin\theta_q\sin\theta_{q'}\sinh\theta_p\right]^2,
\end{align}
where we have used $N_p =\sqrt{m/E_p} $ with the fermionic dispersion relation $E_{p}=\pm \sqrt{p^2 +m^2}$, also we are explicitly using that only the polarization state $\epsilon^{(-)}$ eq.~\eqref{remaining_dr} provides the necessary condition to this Cherenkov decay process to happen.

Furthermore, we can express explicitly the thermal effects in \eqref{eq50} by considering the relations \eqref{F} and \eqref{B}, which allows to obtain
\begin{align}
&\left[\cos\theta_q\cos\theta_{q'}\cosh\theta_p-\sin\theta_q\sin\theta_{q'}\sinh\theta_p\right]^2 = \cr
&=\left(1-n_F(q)\right)\left(1-n_F(q')\right)\left(1-e^{-\beta \omega_p^{(-)}}\right).
\end{align}
Hence, using this result into \eqref{eq50} and also evaluating the trace over the $\gamma$  matrices, we arrive at
\begin{align}
\frac{1}{2}\sum_{spins}|{\cal M}(\beta)|^2=&\,\frac{e^2}{8}\frac{1}{ \omega_p^{(-)2}E_q E_{q-p}}\left(2q^{\mu}q^{\nu}+q^{\nu}p^{\mu}-q^{\mu}p^{\nu}+\frac{1}{2}\eta^{\mu\nu}m_{-}^{2}\right)
\epsilon^{*(-)}_\mu(p)\epsilon^{(-)}_\nu(p)\nonumber\\
&\times \left( \frac{1}{1+e^{-\beta E_q}}\right)  \left( \frac{1}{1+e^{-\beta E_{q-p}}}\right)\left(1-e^{-\beta \omega_p^{(-)}}\right).
\end{align}
This expression can be worked out into its final form.
Thus, after some algebraic manipulations, using the explicit form for the state $\epsilon^{(-)}(p)$ \eqref{remaining_dr} (as well as its normalization condition and orthogonality $p^\mu\epsilon^{(\lambda)}_\mu(p)=0$), we are able to rewrite it as
\begin{align} \label{eq_53}
\frac{1}{2}\sum_{spins}|{\cal M}(\beta)|^2=& \,\frac{e^2}{8}\frac{1}{ \omega_p^{(-)2}E_q E_{q-p}} \left(|\vec{q}|^2\sin^2\theta-\frac{1}{2}m_\gamma^2 +k_{AF}^0|\vec{p}|\right)
\nonumber\\
&\times \left( \frac{1}{1+e^{-\beta E_q}}\right)  \left( \frac{1}{1+e^{-\beta E_{q-p}}}\right) \left(1-e^{-\beta \omega_p^{(-)}}\right),
\end{align}
here $\theta$ is the angle between $\vec{q}$ and $\vec{p}$.

With this development of the transition amplitude \eqref{eq_53} we conclude our analysis in regard of describing the temperature effects within the thermofield dynamics.
The remaining part of the analysis consists in evaluating explicitly the power radiated by the charged particle in terms of \eqref{eq100} and thus determine the thermal effects of the vacuum Cherenkov radiation rate.

\subsection{Radiation rate}
\label{sec5}

Now that we have computed the transition amplitude (and its modulus squared) related with the Cherenkov decay in a thermal bath within the thermofield dynamics, we can finally evaluate the radiation emitted by the charged particle.
It is worth mention that we shall focus our attention in the instantaneous rate of emission, in which the charge emits a single energetic photon, drops below the Cherenkov threshold, and stops emitting \cite{Altschul:2007tn}.

Furthermore, by calculation purposes and to make the reaction kinematics visible, it is convenient to express the integration over the variables $q'$ in \eqref{eq166} as
\begin{equation}
\int\frac{d^{3}q'}{2 q'_0 }=\int d^{4} q'\delta\left( (q'_{0})^2-E_{q'}^{2}\right)\Theta\left(q'_0\right).
\end{equation}
Hence, the total radiation rate at finite temperature $W(\beta)$  eq.~\eqref{eq100}, using the above identity, is cast conveniently as
\begin{align} \label{rad_rate}
W(\beta)  = \frac{m^2}{4\pi^2}\int \frac{d^3p}{ E_q} \delta\left( (q-p)^2-m^2\right) \Theta\left(q_0-p_0\right) \frac{1}{2}\sum_{spins}|{\cal M}(\beta)|^2.
\end{align}
In this case, the energy-momentum conservation for the $e^- \to e^-+\gamma$ decay (shown in the delta function of eq.~\eqref{rad_rate}), implies the following relation
\begin{align} \label{eq310}
m_{\gamma}^{2}-2k_{AF}^0\left|\vec{p}\right|-2\sqrt{\left|\vec{p}\right|^{2}+m_{\gamma}^{2}-2k_{AF}^0\left|\vec{p}\right|}\sqrt{\left|\vec{q}\right|^{2}+m^{2}}+2\left|\vec{q}\right|\left|\vec{p}\right|\cos\theta=0.
\end{align}
One can immediately recognize that the removal of the CPT- and Lorentz-violating effects, through the limit $k^0_{AF} \to 0$, implies $\cos\theta  =1$, which results in a vanishing radiation rate for the Lorentz invariant QED. 

The energy conservation \eqref{eq310} can be used to arrive at the radiation condition for the vacuum decay process.
Hence, from the relation \eqref{eq310} we observe that for finite values of the parameters $\left( k^0_{AF}, m_\gamma^2 ,m^2 \right) \neq 0$ there is a region in the phase space where $\cos\theta <1$, even in vacuum.
This radiation condition corresponds to the availability of a physical phase space for the anomalous decay to happen, and it also corroborates the kinetic condition discussed above.

Furthermore, the energy balance condition requires that the allowed values for the momentum $p$ are only those such that the relation \eqref{eq310} is satisfied for a given value of $\theta$.
This observation corresponds to the fact that the integration over $\theta$ restricts the region of integration over $p$ in \eqref{rad_rate}.
Hence, we conclude that the condition $\cos\theta \in [-1,1]$ restricts the magnitude of the photon momentum $p$ to the values
\begin{align} \label{eq_roots}
p_{\pm } =   \frac{k_{AF}^{0}E_{q}^{2}+\frac{1}{2}m_{\gamma}^{2}\left|\vec{q}\right|-\frac{1}{2}k_{AF}^{0}m_{\gamma}^{2}\pm E_{q}\sqrt{(k_{AF}^{0}E_{q})^{2}-k_{AF}^{0}\left|\vec{q}\right| m_{\gamma}^{2}-m_{\gamma}^{2}m^{2}+\frac{1}{4}m_{\gamma}^{4}}}{  m^{2}+2\left|\vec{q}\right| k_{AF}^{0}-\left(k_{AF}^{0}\right)^{2} }.
\end{align}
It is interesting to observe that the positivity of the square root present in \eqref{eq_roots} implies the condition
\footnote{The other constraint $\left|\vec{q}\right| \leq \frac{m_{\gamma}^{2}-2m\sqrt{m_{\gamma}^{2}-\left(k_{AF}^{0}\right)^{2}}}{k_{AF}^{0}}$ has been ignored, since it becomes negative under the consideration that $m\gg m_\gamma>k_{AF}^{0}$, this follows because $m_\gamma < 10^{-27}\, {\rm GeV} $ and $k_{AF}^{0} \lesssim  10^{-43}\, {\rm GeV} $ \cite{Colladay:2016rmy}.}
\begin{equation}
\left|\vec{q}\right| \geq \frac{m_{\gamma}^{2}+2m\sqrt{m_{\gamma}^{2}-\left(k_{AF}^{0}\right)^{2}}}{2k_{AF}^{0}} \equiv q_{\rm min}.
\end{equation}
This bound is extremely important because it corresponds to the momentum threshold (i.e. $q>q_{\rm min}$) for which the incoming fermion starts to radiate.
Actually, for smaller values, the radiation rate is strictly zero.  

With this development, we can finally evaluate the expression \eqref{rad_rate}.
Hence, making use of spherical coordinates, the total radiation rate \eqref{rad_rate} becomes
\begin{align}
W(\beta)=&\, \frac{e^{2}m^{2}}{32\pi}\frac{1}{\left|\vec{q}\right|E_{q}^2}\int_{p_{-}}^{p_{+}}\frac{p \,dp}{  \omega_{p}^{(-)2} E_{q-p}}\int_{-1}^{1}d\cos\theta\, \delta\left(\cos\theta-\frac{2k_{AF}^{0}\left|\vec{p}\right|+2\omega_{p}^{(-)}E_{q}-m_{\gamma}^{2}}{2\left|\vec{q}\right|\left|\vec{p}\right|}\right)\cr
&\times\left(|\vec{q}|^{2}\sin^{2}\theta-\frac{1}{2}m_{\gamma}^{2}+k_{AF}^{0}|\vec{p}|\right)
\left( \frac{1}{1+e^{-\beta E_q}}\right)  \left( \frac{1}{1+e^{-\beta E_{q-p}}}\right)\left(1-e^{-\beta\omega_{p}^{(-)}}\right)
\end{align}
and by performing the angular integration in $\theta$, we obtain
\begin{align} \label{rad_rate5}
W(\beta)=&\,\frac{e^{2}m^{2}}{32\pi}\frac{1}{\left|\vec{q}\right|E_{q}^{2}} \left( \frac{1}{1+e^{-\beta E_q}} \right) \int_{p_{-}}^{p_{+}}\frac{\,dp}{p\,\omega_{p}^{(-)2}\mathcal{E}_{q-p}}\cr
&\times\left(q^{2}p^{2}-\left(k_{AF}^{0}p+\omega_{p}^{(-)}E_{q}-\frac{1}{2}m_{\gamma}^{2}\right)^{2}-\frac{1}{2}m_{\gamma}^{2}p^{2}+k_{AF}^{0}p^{3}\right) \cr
&\times	\left( \frac{1}{1+e^{-\beta\mathcal{E}_{q-p}}} \right)\left(1-e^{-\beta\omega_{p}^{(-)}}\right),
\end{align}
where we have defined $\mathcal{E}_{q-p}=\sqrt{q^{2}+p^{2}-2k_{AF}^{0}\left|\vec{p}\right|-2\omega_{p}^{(-)}E_{q}+m_{\gamma}^{2}+m^{2}}$.
One straightforward result seen from  \eqref{rad_rate5} is that at the zero temperature limit $T\rightarrow 0 \, (\beta \to \infty)$ the thermal factors go to
\begin{equation}
\left( \frac{1}{1+e^{-\beta E_q}} \right) \left( \frac{1}{1+e^{-\beta\mathcal{E}_{q-p}}} \right)\left(1-e^{-\beta\omega_{p}^{(-)}}\right)\rightarrow 1,
\end{equation}
which corresponds to the standard result at zero temperature.
On the other hand, we observe that the vacuum Cherenkov radiation is modified at very high temperatures.
Interestingly, when the temperature goes to infinity, i.e. $T\to \infty \,(\beta \to 0)$ in \eqref{rad_rate5} the thermal factors now behave as
\begin{equation}
\left( \frac{1}{1+e^{-\beta E_q}} \right) \left( \frac{1}{1+e^{-\beta\mathcal{E}_{q-p}}} \right)\left(1-e^{-\beta\omega_{p}^{(-)}}\right)\rightarrow 0.
\end{equation}
Therefore, there is a (threshold) finite temperature $\beta_{\rm th}$ where the vacuum Cherenkov decay disappears, where the charge simply stops radiating.
This condition can be interpreted as the energy of the thermal bath prohibits the decay to happen even if it is kinematically allowed. \footnote{There are some cases even at zero temperature that the radiation rate is vanishing even then it is allowed by Lorentz-violating kinematics  \cite{Altschul:2017xzx}.}

The integration of \eqref{rad_rate5} is not longer an easy task, as it happens in the zero temperature regime, due to the presence of the thermal distributions.
However, we can evaluate the radiation rate at the high-temperature limit, i.e. $\beta E \ll 1$, and also at some asymptotic regimes in $|\vec{q}|$ which lead to
\begin{align}
W(\beta)\approx\begin{cases}
\frac{e^{2}m^{2}}{64\pi}\frac{\beta^{2}k_{AF}^{0}\left(1+\ln\left(\frac{m^{2}}{2m_{\gamma}^{2}}\right)\right)}{q}, &  q>\frac{m^{2}}{k_{AF}^{0}}\\
\frac{e^{2} }{128\pi} \frac{\beta\left(2k_{AF}^{0}+3m\right)m_{\gamma}}{q}, & q_{\textrm{min}}<q<\frac{m^{2}}{k_{AF}^{0}}.
\end{cases}
\end{align}
We can immediately observe that both regimes give a temperature dependent radiation rate: the higher energy rate (at $q>\frac{m^{2}}{k_{AF}^{0}}$) decreases faster as the temperature increases due to the $\beta^2$ factor, until it goes to zero at the threshold temperature $\beta_{\rm th}$.
It is curious to notice that the thermal effects have changed the momentum $q$ dependence of the radiation rate: we have found that at finite temperature both behave as $1/q$, while at zero temperature the rate at the low-energy and high-energy regime have quadratic and linear dependence on $q$, respectively 
\cite{Colladay:2016rmy}.

One last remark that we would like to present is about the case of spacelike $k_{AF}^{\mu}$.
Actually, the main difference of the present analysis with the spacelike $k_{AF}^{\mu}$ case are the dispersion relations and related polarization states.
Obviously, the radiation rate at the high-temperature limit will be significantly modified, but our main conclusion still holds: there is a finite temperature where the particle simply stops emitting radiation.

Hence, we believe that the present analysis we found interesting results and conclusions, elucidating some aspects of the anomalous decays at finite temperature, and that the spacelike $k_{AF}^{\mu}$ case would not provide any other physically relevant conclusion in addition to those discussed here.

\section{Final remarks}
\label{conc}

In this work we have studied the emission of the vacuum Cherenkov radiation at finite temperature. 
Since anomalous decay processes possess unique signature of Lorentz violation, they are a suitable scenario to study phenomena beyond the standard model.
In particular, it is expected that the behavior of these novel phenomena can drastically change in the presence of a heat bath: one can naively expect that it can enhance the effects or even prohibits them to happen.
Hence, we have examined the rate of radiated energy from a charged particle through vacuum Cherenkov within the thermofield dynamics approach.

In our analysis of the radiation emission at finite temperature we have considered a model of the minimal sector of the SME, where this decay process is known to happen at $T=0$, which corresponds to the CPT- and Lorentz-violating photon  Lagrangian. 
We have revised the main aspects of this model, focusing in obtain its dispersion relation and the respective polarization vectors.
In particular, only one mode $\omega_p^{(-)}$ and $\epsilon^{(-)}_{\mu}$ of the photon field is responsible to engender instabilities in its propagation such that its phase velocity is subluminal, and thus satisfying the kinematic constraint related with the radiation emission.

Since the thermofield dynamics formalism allows to compute transition amplitudes exactly as in the $T=0$ field theory, by doubling the set of operators and its respective Fock space (where the second set acts like a heat bath), we have explicitly evaluated the scattering matrix element related with the process $e^- \to e^- + \gamma$, showing in details how the temperature effects are incorporated.
After some algebraic steps, we arrived at the expression of the radiation rate \eqref{rad_rate5}. Due to the presence of the thermal distributions the remaining integration does not possess a closed form, hence we evaluate it at some limits of interest: 
\begin{itemize}

\item In the high-temperature regime $\beta E \ll 1$ we have found that the thermal effects have changed the momentum $q$ dependence of the radiation rate (in comparison with the zero temperature case);

\item The higher energy expression (at $q>\frac{m^{2}}{k_{AF}^{0}}$) is more sensitive to the temperature effects, due to the presence of the $\beta^2$ factor, decreasing faster as the temperature increases, until it goes to zero at the threshold temperature $\beta_{\rm th}$;

\item Interestingly, one can straightforwardly observe from  \eqref{rad_rate5} that at the limit  $T\to \infty \,(\beta \to 0)$, the radiation rate expression goes to zero: this can physically be interpreted as if the energy of the thermal bath (at some temperature threshold $\beta_{\rm th}$) prohibits the decay to happen even if it is kinematically allowed.

\end{itemize}

In summary, we have explicitly shown how thermal effects can change the outcome of physical effects in a anomalous decay related with physics beyond the standard model.
Certainly, there are a number of anomalous phenomena that can have their known behavior modified by the presence of a heat bath.
Since tiny changes in electromagnetic wave propagation can be
scrutinized from TeV photons data (from extremely energetic
astronomical sources), presenting itself as a phenomenologically rich environment, we shall focus in this sector to examine further examples where the temperature can modify significantly the dynamics of the model.

 \subsection*{Acknowledgements}

R.B. acknowledges partial support from Conselho
Nacional de Desenvolvimento Cient\'ifico e Tecnol\'ogico (CNPq Projects No. 305427/2019-9) and Funda\c{c}\~ao de
Amparo \`a Pesquisa do Estado de Minas Gerais (FAPEMIG Project No. APQ-01142-17). The work by A. F. S. is partially supported by Conselho
Nacional de Desenvolvimento Cient\'ifico e Tecnol\'ogico (CNPq Project No. 313400/2020-2). 



\global\long\def\link#1#2{\href{http://eudml.org/#1}{#2}}
 \global\long\def\doi#1#2{\href{http://dx.doi.org/#1}{#2}}
 \global\long\def\arXiv#1#2{\href{http://arxiv.org/abs/#1}{arXiv:#1 [#2]}}
 \global\long\def\arXivOld#1{\href{http://arxiv.org/abs/#1}{arXiv:#1}}


\end{document}